\documentclass[aps,prl,twocolumn,superscriptaddress]{revtex4-1}
\usepackage{graphicx,amsmath,amssymb,bm}
\usepackage[
   bookmarks=true,
   colorlinks=true,
   hyperindex=true,
   citecolor=blue,
   linkcolor=blue,
   urlcolor=red
]{hyperref}

\begin{document}

\title{Observation of Crossover from Repulsive to Attractive Interactions between Two $nS$-Rydberg Atoms in Dual-Species Atom Arrays}

\author{Yi Chen}
\affiliation{Innovation Academy for Precision Measurement Science and Technology, Chinese Academy of Sciences, Wuhan 430071, China}
\affiliation{School of Physical Sciences, University of Chinese Academy of Sciences, Beijing 100049, China}

\author{Kun-Peng Wang}
\email{wangkunpeng@wipm.ac.cn}
\affiliation{Innovation Academy for Precision Measurement Science and Technology, Chinese Academy of Sciences, Wuhan 430071, China}

\author{Ming-Rui Wei}
\affiliation{Innovation Academy for	Precision Measurement Science and Technology, Chinese Academy of Sciences, Wuhan 430071, China}
\affiliation{School of Physical Sciences, University of Chinese Academy of Sciences, Beijing 100049, China}

\author{Peng Zhang}
\affiliation{Innovation Academy for	Precision Measurement Science and Technology, Chinese Academy of Sciences, Wuhan 430071, China}
\affiliation{School of Physical Sciences, University of Chinese Academy of Sciences, Beijing 100049, China}

\author{Yang Wang}
\affiliation{Innovation Academy for	Precision Measurement Science and Technology, Chinese Academy of Sciences, Wuhan 430071, China}
\affiliation{School of Physical Sciences, University of Chinese Academy of Sciences, Beijing 100049, China}

\author{Yi-Bo Wang}
\affiliation{Innovation Academy for Precision Measurement Science and Technology, Chinese Academy of Sciences, Wuhan 430071, China}

\author{Min Liu}
\affiliation{Innovation Academy for Precision Measurement Science and Technology, Chinese Academy of Sciences, Wuhan 430071, China}

\author{Peng Xu}
\affiliation{Innovation Academy for	Precision Measurement Science and Technology, Chinese Academy of Sciences, Wuhan 430071, China}
\affiliation{Wuhan Institute of Quantum Technology, Wuhan 430206, China}

\author{Xue-Yuan Wang}
\affiliation{National Key Laboratory of Computational Physics, Institute of Applied Physics and Computational Mathematics, Beijing 100088, China}

\author{Xiang Gao}
\affiliation{National Key Laboratory of Computational Physics, Institute of Applied Physics and Computational Mathematics, Beijing 100088, China}

\author{Chengshu Li}
\affiliation{Chen-Ning Yang Institute for Advanced Study, Tsinghua University, Beijing, 100084, China}
\affiliation{Beijing Key Laboratory of Cold Atom Quantum Computation, Tsinghua University, 100084, Beijing, China}

\author{Xiao-Dong He}
\email{hexd@wipm.ac.cn}
\affiliation{Innovation Academy for Precision Measurement Science and Technology, Chinese Academy of Sciences, Wuhan 430071, China}
\affiliation{Wuhan Institute of Quantum Technology, Wuhan 430206, China}

\author{Ming-Sheng Zhan}
\email{mszhan@apm.ac.cn}
\affiliation{Innovation Academy for Precision Measurement Science and Technology, Chinese Academy of Sciences, Wuhan 430071, China}
\affiliation{Wuhan Institute of Quantum Technology, Wuhan 430206, China}
\affiliation{Hefei National Laboratory, Hefei 230088, China}

\date{\today}

\begin{abstract}
Dual-species Rydberg atom arrays provide a flexible platform for engineering long-range interactions beyond the conventional isotropic blockade regime. We spectroscopically study interactions between different \(s\)-state Rydberg levels in a \(^{87}\mathrm{Rb}\)--\(^{85}\mathrm{Rb}\) optical-tweezer array. By varying the principal-quantum-number difference $\Delta n$ around $61S_{1/2}$, we observe a crossover from repulsive to attractive interactions. We identify a distinct regime at \(\Delta n=2\) in which F\"orster-enhanced electronic spin exchange dominates over orbital exchange and produces a strongly bond-orientation-dependent pair spectrum. Time-domain collective-excitation measurements reveal this geometry-dependent spin exchange through a pronounced anisotropy of the double-excitation probability. For \(\Delta n=1\), by contrast, orbital exchange dominates and the interaction is nearly isotropic. These results establish the principal-quantum-number difference and bond orientation as complementary control parameters for tailoring Rydberg interaction spectra and collective excitation dynamics, opening new opportunities for engineering spin-dependent Hamiltonians in programmable atom arrays.
\end{abstract}
\maketitle

\emph{Introduction}---
Rydberg atoms in optical-tweezer arrays have emerged as a powerful platform for quantum computation \cite{Saffman2026,HZhou2026} and quantum simulation \cite{Browaeys2020}.
In the widely used regime, atoms are excited to the same \(nS\) Rydberg state and interact through repulsive van der Waals interactions arising from second-order dipole--dipole couplings \cite{Lukin2000,Saffman2010}. The resulting nearly isotropic Rydberg blockade provides local constraints that underpin high-fidelity entanglement \cite{Lukin2026}, effective PXP models \cite{Papic2018,Lukinscars2019,Papic2021,YCheng2024}, and a broad range of many-body phenomena \cite{Lukin2017,Lukin2019,Dalmonte2020,LGT2025,LukinQSL2021,Browaeys2025}. Rydberg interactions, however, are richer than a simple repulsive potential. Extending this interaction toolbox to include controllable attractive interactions and spin-dependent exchange processes offers new opportunities for engineering quantum many-body models, tuning quantum criticality, and exploring emergent quantum phases~\cite{Lan2016,Alicea2021,SUSY2024,SUSY2025,Liu2026}.

The richness of Rydberg interactions arises from the interplay of virtual dipole couplings, energy defects, and exchange processes, which jointly determine their sign, magnitude, and angular dependence  \cite{Saffman2008,Saffman2010}. In particular, preparing two atoms in different \(s\)-Rydberg levels provides a direct route to tuning the interactions and even reversing their sign \cite{Cano2012,XFShi2014,Beterov2015,Mandoki2016,CSAdams2025,LiZhai2026}.
Microscopically, such pairs can support two distinct exchange processes: orbital exchange, in which the atoms exchange their principal quantum numbers, and electronic spin exchange, in which they exchange their spin projections. These processes exhibit different dependences on the principal-quantum-number difference $\Delta n$, fine structure, and bond orientation, and their interplay is closely related to F\"orster resonances and anisotropic spin exchange  \cite{Cano2012,XFShi2014,Beterov2015,Mandoki2016,CSAdams2025,LiZhai2026}, and lead to spin-position locked phenomena \cite{LiZhai2026}.

Dual-species arrays \cite{Sheng2022,Bernien2022,KKNi2022,Takahashi2025,Saffman2025} provide a versatile platform for exploring these interactions, as the two species can be independently prepared, spectrally addressed, and excited to different Rydberg states \cite{Anand2024,SaffmanRbCs2026,KKNi2026}. This independent control offers additional flexibility in engineering interspecies and intraspecies interactions and designing many-body dynamics \cite{Chepiga2024,Tomza2025,Pichler2026}.  In particular, this independent control allows $\Delta n$ to be varied while maintaining fixed target spin projections, facilitating the identification of orbital- and spin-exchange contributions.
Although coherent orbital- and spin-exchange dynamics have been demonstrated in single-isotope  systems \cite{Browaeys2025exchange,CChen2026}, how these competing exchange processes shape the orientation-dependent interaction spectrum and govern collective Rydberg excitation remains insufficiently explored. Establishing this connection is important for engineering spin-dependent interactions \cite{LiZhai2026} in programmable atom arrays and exploring nonequilibrium many-body dynamics \cite{Lukin2019,SUSY2025,LiZhai2026}.

\begin{figure*}[htbp]
	\centering
	\includegraphics[width=15cm]{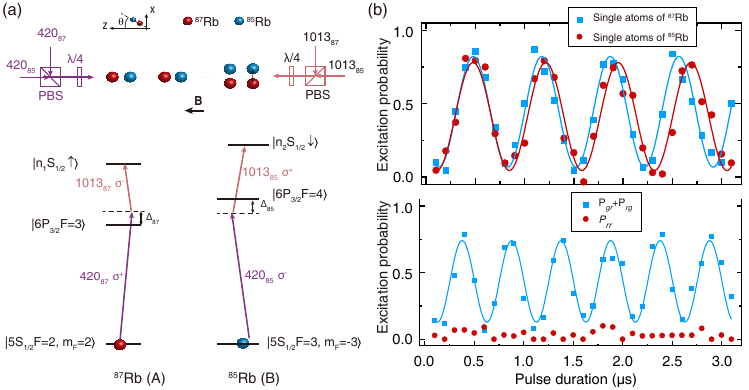}
	\caption{Dual-species Rydberg excitation of $^{87}$Rb and $^{85}$Rb. (a) Schematic diagram of dual-species Rydberg excitation and relevant energy levels. $^{87}$Rb atoms are excited to $|n_1S\rangle = |61S\rangle$ state with $m_J = 1/2$ denoted by $|\uparrow\rangle$. $^{85}$Rb atoms are excited to $|n_2S\rangle$, ranging from $|59S\rangle$ to $|65S\rangle$, with $m_J = -1/2$ denoted by $|\downarrow\rangle$. The excitations are driven by two-photon transitions with 420-nm and 1013-nm lasers both with linewidths of less than 1 kHz. (b) Coherent Rabi oscillations of single atoms (upper panel, $^{85}$Rb:  $|59S\rangle$) and collective excitation of single-species atom pairs (lower panel, Atom \(B\)). The collective oscillation frequency of $P_{gr}+P_{rg}$ is enhanced by $\sqrt{2}$ relative to the single-atom Rabi frequency. }
	\label{fig1}
\end{figure*}

Here we investigate Rydberg interactions in a dual-isotope $^{87}$Rb–$^{85}$Rb optical-tweezer array using interaction spectroscopy and collective-excitation measurements. By varying the principal-quantum-number difference $\Delta n$ around $61 S_{1/2}$, we observe a crossover from repulsive to attractive interactions and identify distinct orbital- and spin-exchange-dominated regimes.
In particular, for $\Delta n=2$, F\"orster-enhanced electronic spin exchange produces a strongly orientation-dependent interaction spectrum, with a near-resonant branch appearing for parallel alignment. This spectral anisotropy is reflected in the collective-excitation dynamics, where rotating the interatomic axis switches the response between appreciable double excitation and strong Rydberg blockade. Our results establish principal-quantum-number and geometric control of spin-dependent Rydberg interactions, providing a route toward programmable many-body Hamiltonians.

\emph{Experimental Setup}---
The experimental setup was described previously in Refs. \cite{setup2019,Sheng2022,array2026}, and the scheme for dual-species Rydberg excitation is depicted Fig.\ref{fig1}(a).
We prepare defect-free dual-species arrays by rearrangment from a static $ 6 \times 26 $ tweezer array.  The tweezer trap spacings are tuned between \( 4.88(6)~\mu\)m and \( 6.65(6)~\mu\)m.
A bias magnetic field $B_z = 7.17$ G defines the quantization axis \(z\). \(^{87}\)Rb (atom \(A\)) is prepared in the ground state \(|g_A\rangle \equiv |5S_{1/2}, F=2, m_F=+2\rangle\) and \(^{85}\)Rb (atom \(B\)) in \(|g_B\rangle \equiv |5S_{1/2}, F=3, m_F=-3\rangle\). These two states have opposite electronic-spin projections, thus, they can be coupled to Rydberg states in opposite $m_J$ state by two-photon excitations. The two-photon excitation is species-selective: the 420-nm laser that couples ground $|5S_{1/2}\rangle$ and intermediate $|6P_{3/2}\rangle$ states is $\sigma^+$ ($\sigma^-$) polarized and detuned by about $630$ MHz ($-600$ MHz), for atom \(A\) (\(B\)), while the 1013-nm laser that couples $|6P_{3/2}\rangle$ and Rydberg levels is $\sigma^-$ ($\sigma^+$) polarized respectively. This drives atom \(A\) to
$|n_1 S_{1/2}, m_J=+1/2\rangle_A \equiv |n_1 S,\uparrow\rangle $ and atom \(B\) to
$|n_2 S_{1/2}, m_J=-1/2\rangle_B \equiv |n_2 S,\downarrow\rangle$.
The atoms are arranged that the interatomic axis is oriented at an angle \(\theta\) with respect to the quantization axis $z$. Single-atom calibrations and the $\sqrt{2}$ collective Rabi enhancement for same-species pairs establish coherent excitation, as shown in  Fig. \ref{fig1}(b).
Collective excitation is monitored by preparing two closely spaced atoms of $4.88 \mu $m, and measuring the probability of single excitation \(P_{gr}\) + \(P_{rg}\)  that one atom is in Rydberg states. Because the lasers are fixed at the single-atom resonances, the double-Rydberg state \(P_{rr}\) is suppressed due to Rydberg interactions in the same $s$-state and only coupled when the interaction energy \(V_{\mathrm{int}}\) is small compared to two-photon Rabi frequency \cite{Browaeys2009}. We vary the second principal quantum number $n_2$ from 59 to 65 while holding $n_1=61$, and compare interaction spectra with different geometries and spacing.

\begin{figure}[htbp]
	\centering
	\includegraphics[width=8.5cm]{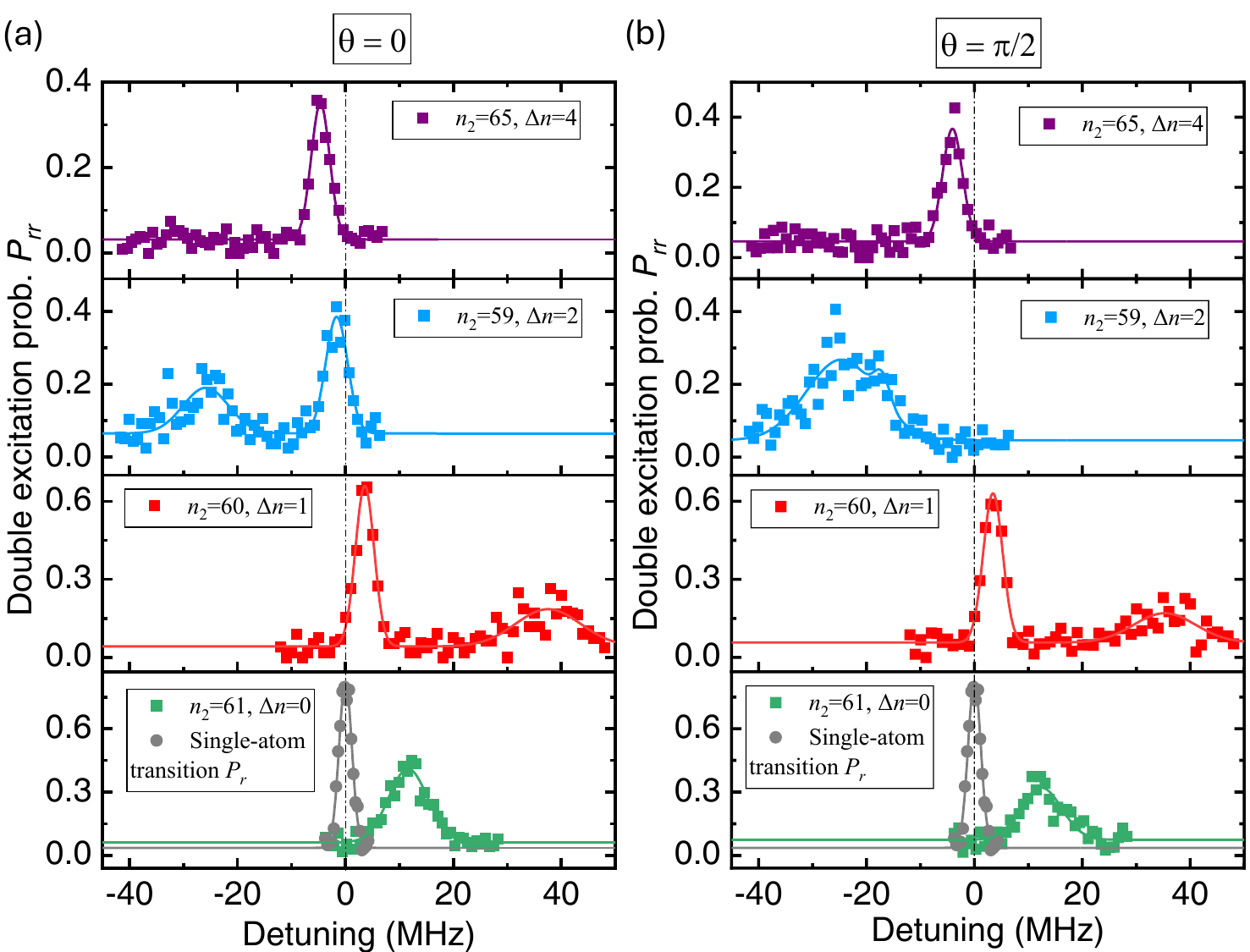}
	\caption{Two-body Rydberg interaction spectra in a dual-species $^{87}$Rb-$^{85}$Rb optical-tweezer array with a spacing of 4.88 $\mu$m.  Measured double-excitation probability \(P_{rr}\) versus the two-photon detuning of the atom \(B\) excitation for $\Delta n= 1-4$, where $\Delta n = |n_2-n_1|$, $n_1=61$. Solid lines are Gaussian-fitting to the data. Detunings are relative to corresponding single atom transition frequencies (gray circles, only $n_2 = 59$ was shown). (a) The atoms are arranged in the parallel configuration \(\theta=0\) (atoms aligned parallel to $z$).  The fitted peaks  are $\Delta n=0$: 11.6(2) MHz, $\Delta n=1$: [3.58(9), 37.5(8)] MHz, $\Delta n=2$: [-25.9(6), -1.6(2)] MHz, $\Delta n=4$: -4.63(1) MHz. (b) Perpendicular configuration \(\theta=\pi/2\). The fitted peaks  are $\Delta n=0$: 12.4(3) MHz, $\Delta n=1$: [3.44(9), 35.3(8)] MHz, $\Delta n=2$: [-24.6(7), -17.1(6)] MHz, $\Delta n=4$: -4.1(1) MHz. }
	\label{fig2}
\end{figure}

\begin{figure}[htbp]
	\centering
	\includegraphics[width=8.5cm]{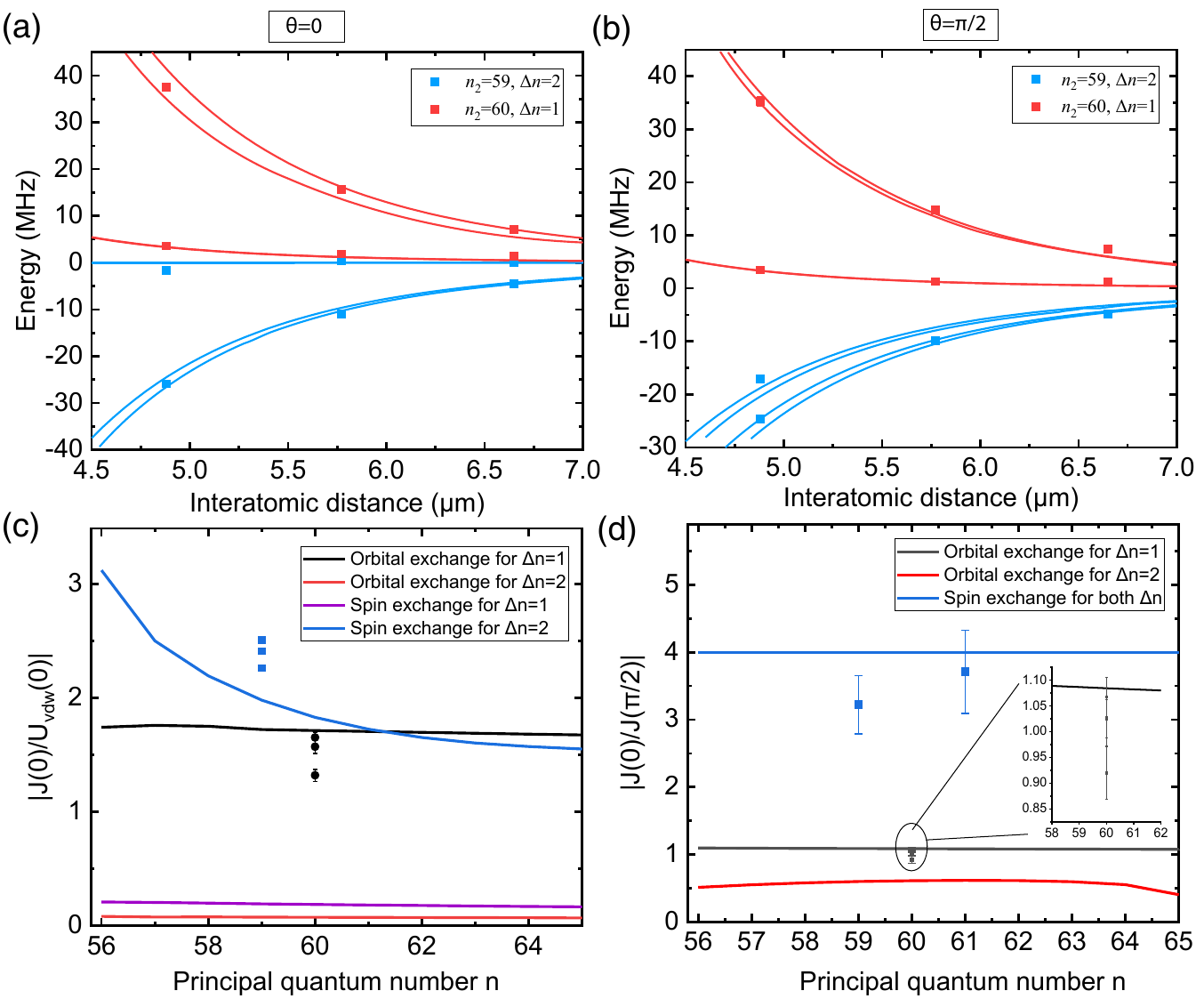}
	\caption{Measured interaction energies and exchange coefficients (Solid lines are \textit{ab initio} calculations). Measured interaction energies for $\Delta n = 1$  and 2 are extracted by fitting spectra at three distances for (a) $\theta= 0$ and (b) $\theta= \pi/2$. (c) and (d), orbital- and spin- exchange coefficients for $|n S\rangle$ + $|(n + \Delta n) S\rangle$ pairs, which are calculated from measured interaction energies in (a) and (b).  (c) The exchange strength  $\theta= 0$ is characterized by the ratio to diagonal van der Waals interactions, $J(0)/U_{vdW}(0)$. Spin exchange for $\Delta n= 1$ and Orbital exchange for  $\Delta n = 2$ are very weak. Data points are experimental values at three distances extracted from (a), Squares (Filled circles) represent $\Delta n= 1$ (2). (d) Anisotropy is characterized by the ratio of $J_{xy}$ in the $\theta= 0$  and $\pi/2$ directions. Data points are experimental values  at $4.88 \mu$m. }
	\label{fig3}
\end{figure}

\emph{Interaction spectroscopy}---
We spectroscopically measure interactions between $n_1 S$-$n_2 S$ states by first exciting atom \(A\) with a resonant $\pi$ pulse, then scanning the detuning of atom \(B\) and probing the double excitation probabilities \(P_{rr}\) for the \(A\)-\(B\) pairs. This sequential protocol separates the pair-interaction shift from the single-particle optical resonance. For an interaction eigenstate of shift $U$, the second excitation is favored when its detuning compensates $U/h$; a positive (negative) resonance shift therefore indicates a repulsive (attractive) branch. Because the laser couples to a superposition of pair eigenstates, the spectrum contains both energy and  optical overlap  rather than a single scalar interaction coefficient. We study two configurations: parallel configuration, \(\theta=0\) (atoms aligned parallel to $z$) and perpendicular configuration \(\theta=\pi/2\). As shown in Fig.\ref{fig2}, when varying $\Delta n$ from 0 to 4, we observe a crossover from repulsive to attractive interactions. At $\Delta n=0$, double excitation is shifted to the repulsive side and is strongly blockaded close to the single-atom resonance. At $\Delta n=4$, the prominent feature moves to negative detuning, indicating attractive interactions. For these two cases, the measured interaction energy is in accordance with $-C_6/R^6$, where $C_6$ is the calculated van der Waals coefficients with PairInteraction or ARC package \cite{Pairint2026,ARC}. Between these limits, the $\Delta n=1$ and $2$ manifolds show richer structures, including branches near the bare single-atom resonance. The central distinction is geometric: for $\Delta n=1$ the response changes comparatively little between the two bond orientations, whereas at $\Delta n=2$ a near-resonant branch is visible for $\theta=0$ but the perpendicular configuration is strongly shifted and suppresses resonant double excitation. Thus the change of $\Delta n$ controls not only the sign but also the spin selectivity and angular contrast of the pair spectrum.
We then measure the interaction spectra at distinct interatomic distances and extract interaction peaks as shown in Fig. \ref{fig3}(a) and \ref{fig3}(b)  (Spectrum for $n_2$ = 61 was not shown in Fig. \ref{fig2}). The measured values are comparable with \textit{ab initio} calculations 
The calculations are performed in full two-atom basis that contains pair-state combinations regardless spin projections and its exchanges. Thus full spectra exhibit minor splittings, e.g. orbital (spin) -exchange effect in $\Delta n$  = 2 (1), which were not clearly resolved from measured spectra. Second-order approximation can be applied to get effective Hamiltonians to study the microscopic origin and dynamics below. 

\emph{Microscopic origin}---For an interatomic displacement ${\bf R}$, the electric dipole interaction is $V_{dd}=[{\bf d}_A\cdot{\bf d}_B-3({\bf d}_A\cdot\hat{\bf R})({\bf d}_B\cdot\hat{\bf R})]/(4\pi\epsilon_0R^3)$ \cite{Saffman2010}. Its second-order coupling through nearby $P_{1/2}$ and $P_{3/2}$ pair manifolds generates the effective $S$--$S$ interaction. For atom \(A\)-\(B\) pairs in different Rydberg $s$-states, except for diagonal van der Waals interactions, they can exchange their principal quantum number (orbit) through virtual \(P-P\) channels, as observed in Ref. \cite{Browaeys2025exchange}, and resulting a XXZ Hamiltonian with matrix element in opposite-state subspace of
$H_{orb}=\begin{pmatrix}U_{vdW} & J_{orb}/2\\
J_{orb}/2&U_{vdW}\end{pmatrix}$ regardless of interaction angle \cite{Weidemuller2021}.
In addition, for two atoms in opposite $m_J$ states, they can in principle exchange their electronic-spin projection directly and resulting a orientation $\theta$-dependent Hamiltonian in opposite-spin subspace of
$H_{spin}=\begin{pmatrix}U_{vdW}(\theta)-J_z(\theta)/4 & J_{xy}(\theta)/2\\
	J_{xy}(\theta)/2&U_{vdW}(\theta)-J_z(\theta)/4\end{pmatrix}$.
To understand the underlying difference for the spectrum with $\Delta n$  = 1 and 2, we perform \textit{ab initio} calculations \cite{Pairint2026} of the orbital- and spin-exchange coefficients and compare their strengths and anisotropy with measured values, as shown in Fig.\ref{fig3}(c) and Fig.\ref{fig3}(d). Orbital exchange coefficient is calculated by $J_{orb} =2 \langle n S, \uparrow; (n+\Delta n) S, \downarrow| H |(n+\Delta n) S, \uparrow; n S, \downarrow\rangle$ and spin-exchange coefficients  $J_{xy} =2\langle n S, \uparrow; (n+\Delta n) S, \downarrow| H | n S, \downarrow; (n+\Delta n) S, \uparrow\rangle$. Calculations show that the orbital exchange is dominant and nearly isotropic for $\Delta n=1$; the electronic-spin exchange is enhanced and anisotropic for $\Delta n=2$. Coefficients for other combinations are at least one order of magnitude smaller.
Orbital-exchange for $\Delta n = 1$  is strong and nearly isotropic, $J_{orb}$ is nearly the same for $\theta= 0$ and $\pi/2$. The corresponding experimental values are obtained by solving equations of the measured peaks with the eigenvalues of matrix $H_{orb}$ 
. The isotropy is a direct reflection of the interaction spectrum in Fig.\ref{fig2}. Spin-exchange has a universal ratio of $J_{xy}(0)/J_{xy}(\pi/2)=4$ that determined by angular-momentum coupling coefficients. For $\Delta n = 2$, spin exchange is F\"orster enhanced compared with other $\Delta n$ values due to electronic spin-orbit coupling of underlying interaction channels, that contain both $|P_{1/2}\rangle$ and $|P_{3/2}\rangle$ states and resulting in a spin-dependent interaction.
The resulting exchange term in $H_{spin}$ is close to diagonal elements for \(\theta=0\), leading to a near zero eigenvalue, whereas for \(\theta=\pi/2\), exchange term is much smaller than diagonal elements.
The experimental $J_{xy}$ ratios are obtained by solving equations of the measured peaks with the eigenvalues of matrix $H_{spin}$ with the relation $J_z(0)/J_{xy}(0)=-1/2$ and $J_z(\pi/2)/J_{xy}(\pi/2)=4$.
The strong deviation from an isotropic ratio of 1 provides spectroscopic evidence for electronic-spin exchange when $\Delta n = 2$.

\begin{figure}[htbp]
	\centering
	\includegraphics[width=8.5cm]{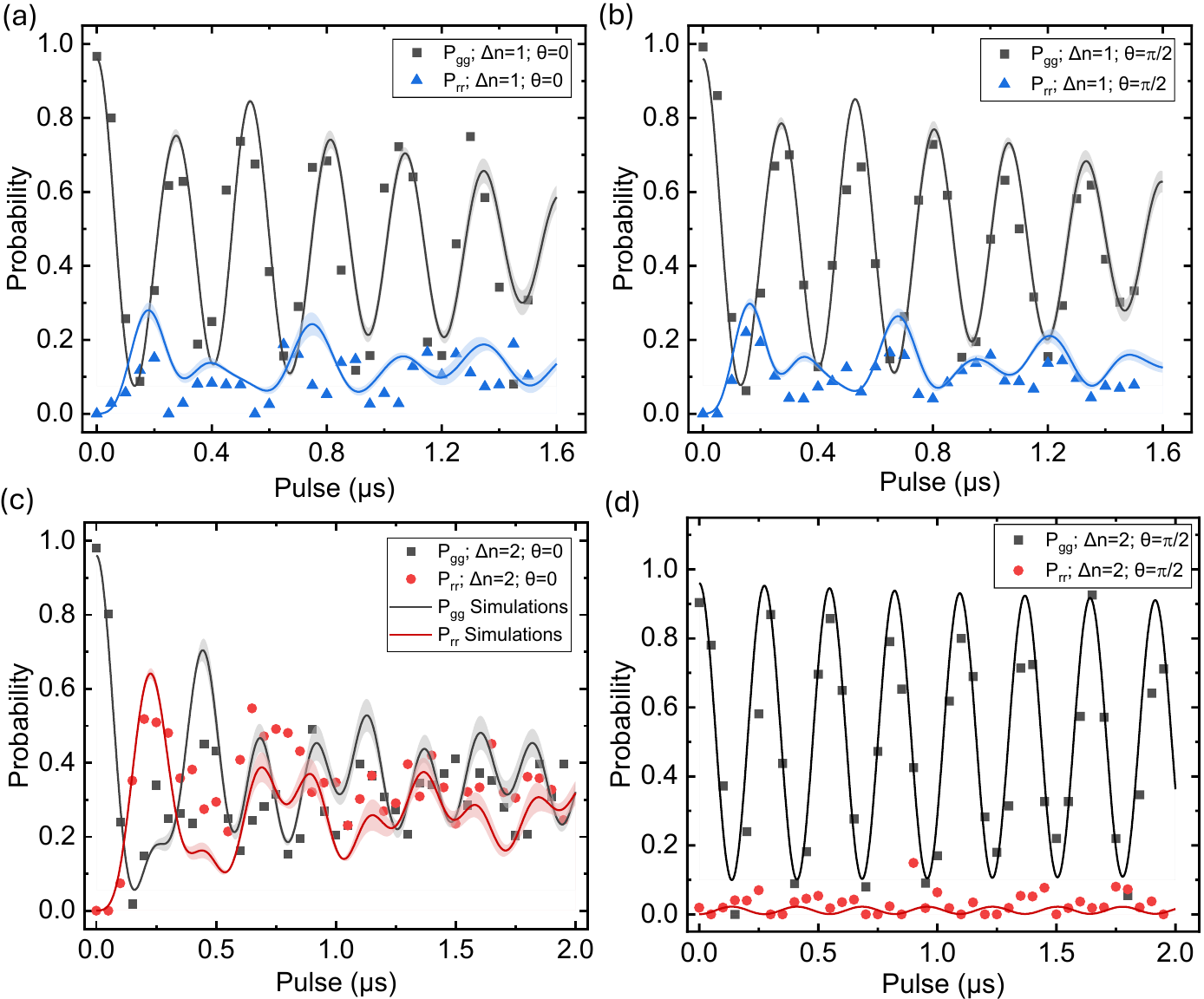}
	\caption{Collective excitation dynamics of dual-species \(A\)-\(B\) pairs with spacing of $4.88$ $\mu$m. Solid lines are simulations with experimental parameters for each $\Delta n$. Rabi couplings are tuned to be balanced. (a) and (b), $n_2 = 60$ and $\Delta n$ = 1. $\Omega_A=\Omega_B=2\pi*2.87$ MHz. $P_{rr}$ oscillations have similar behavior for $\theta = 0$ and $\pi/2$. (c) and (d), $n_2 = 59$ and $\Delta n$ = 2. $\Omega_A=\Omega_B=2\pi*2.56$ MHz.  For \(\theta=0\), significant double excitation probability \(P_{rr}\) is observed. And for \(\theta=\pi/2\), \(P_{rr}\) is suppressed,  \(P_{gg}\) oscillates at a frequency of $\sqrt{2}$ times the single-atom Rabi frequency. }
	\label{fig4}
\end{figure}

\emph{Dual-species collective excitation}---To further characterize the dual-species interaction, we prepare \(A\)-\(B\) pairs and drive both species simultaneously at their single-atom resonances and observe the collective excitation dynamics. The time-domain measurement  probes physics that a static interaction spectrum cannot resolve. In the absence of interactions, the two drives populate the doubly excited state through the two single-excitation pathways. A large interaction shift detunes the final step and yields blockade, whereas a near-resonant eigenbranch restores a coherent pathway to double excitation.
As shown in Fig. \ref{fig4}, the near-zero branch leads to a high \(P_{rr}\) probability compared with conventional single-species collective excitation. While the dynamics shows nearly isotropic for $\Delta n =1$, it is strongly geometry dependent for $\Delta n =2$.  For $\Delta n =2$ in the perpendicular configuration (\(\theta=90^\circ\)), where no near-zero branch exists, the \(P_{rr}\) is strongly suppressed as  in conventional single-species Rydberg blockade. This qualitative trend is the experimental signature of the spin-position-coupled Rydberg interaction in the dual-species system.

To quantify the collective excitation dynamics, we perform numerical simulations with the master equation $d\rho/dt=-i[H,\rho] + L[\rho]$ 
. We study the two-atom system in a Hilbert space spanned by the internal states \(|s_A\rangle \otimes |s_B\rangle\), where \(s_i \in \{g_i, r_i\}\) and the Rydberg states carry spin \(m_J=\pm 1/2\). The initial state is
\(|\psi(0)\rangle = |g_A\rangle \otimes |g_B\rangle\).
The probability of double Rydberg excitation is evaluated as
$P_{rr}(t) = \langle n_1 n_2 \rangle(t)$. The Hamiltonian in the rotating frame reads
$H = H_{\mathrm{laser}} + H_{\mathrm{det}} + H_{\mathrm{Zeeman}} + H_{\mathrm{int}}$.
The laser term is spin-selective:
$H_{\mathrm{laser}} = \Omega_A/2\left(|n_1 S,\uparrow\rangle\langle g_A| + \mathrm{h.c.}\right)
+ \Omega_B/2\left(|n_2 S,\downarrow\rangle\langle g_B|
+ \mathrm{h.c.}\right)
$,
with Rabi frequencies \(\Omega_A\) and \(\Omega_B\). The detuning term is
$H_{\mathrm{det}} = \delta_A N_A + \delta_B N_B$, with $N_i = |r_i\rangle\langle r_i|  (i=A,B)$,
and we set \(\delta_A=\delta_B=0\) to represent lasers tuned to single-atom resonances. $H_{\mathrm{Zeeman}}$ is the Zeeman term.
The interaction term acts only when both atoms are in Rydberg states: $H_{\mathrm{int}} = n_A n_B \, H_{\mathrm{eff}}$,
where \(H_{\mathrm{eff}}\) is the effective spin-interaction Hamiltonian in the projected subspace of orbital exchange for $\Delta n=1$ ($H_{orb}$) and spin exchange  for $\Delta n=2$ ($H_{spin}$) and the corresponding matrix elements are calculated numerically 
.
For each $\Delta n$ and angle \(\theta\), we calculate the interaction matrix accordingly as input parameters. And dissipation term in the master equation is added especially the interaction spectrum broadening 
.
The simulation qualitatively captures the experimental observation in interaction spectra.
For $\Delta n =2$, \(P_{rr}\) is substantially larger for \(\theta=0\) than for \(\theta=\pi/2\). This confirms that the near-zero interaction branch in the parallel configuration makes the double-Rydberg state near resonant with the single-atom transitions, thereby enhancing collective excitation, while the perpendicular configuration remains strongly blockaded.

In summary, we have observed geometry-dependent two-body Rydberg interactions in a dual-species $^{87}$Rb-$^{85}$Rb optical-tweezer array. By varying the principal-quantum-number difference between two $s$-state Rydberg atoms around $61S_{1/2}$, we observe a crossover from repulsive to attractive interactions. For $\Delta n$ = 2, the interaction becomes strongly orientation-dependent due to enhanced electronic-spin exchange, which has distinct spectra and collective excitation behavior. These observations motivate geometry-programmable interactions for dual-species Rydberg arrays, and provide a versatile resource for engineering many-body Hamiltonians \cite{SUSY2024,SUSY2025,LiZhai2026} and quantum gates \cite{XFShi2014,Beterov2015}.

\emph{Acknowledgments}---
We thank Hui Zhai for discussions.
This work is supported by the National Natural Science Foundation of China under grants No. 12241410, 12374251, U22A20257, 12504307, the National Key Research and Development Program of China under Grant No. 2021YFA1402001, the National Innovation Program for Quantum Science and Technology of China under Grant No. 2023ZD0300401, 2025ZD0300400, the Strategic Priority Research Program of the Chinese Academy of Sciences XDB1690202, XDB1690201, the Major Program (JD) of Hubei Province under Grant No. 2023BAA020. C.L. is also supported by the Tsinghua University Dushi program.




\begin{thebibliography}{99}

\bibitem{Saffman2026}
M. Saffman, Neutral atom quantum computing, arXiv: 2608.30783 (2026).

\bibitem{HZhou2026}
A. J. Menssen \textit{et al}, Strategic Plan for Neutral Atom Quantum Computation, arXiv: 2607.21554 (2026).

\bibitem{Browaeys2020} A. Browaeys and T. Lahaye, Many-body physics with individually controlled Rydberg atoms. Nat. Phys. \textbf{16}, 132-142 (2020).

\bibitem{Lukin2000}
D. Jaksch, J. I. Cirac, P. Zoller, S. L. Rolston, R. C\^{o}t\'{e}, and M. D. Lukin, Fast Quantum Gates for Neutral Atoms, Phys. Rev. Lett. \textbf{85}, 2208 (2000).

\bibitem{Saffman2010}
M. Saffman, T. G. Walker, and K. M\o{}lmer, Quantum information with Rydberg atoms, Rev. Mod. Phys. \textbf{82}, 2313 (2010).

\bibitem{Lukin2026}
S. J. Evered, M. Xu, S. H. Li, A. A. Geim, J. P. B. Ataides, M. Kalinowski, D. Bluvstein, N. Maskara, C. Kokail, M. Greiner, V. Vuleti\'c, and M. D. Lukin, High-fidelity entangling gates and nonlocal circuits with neutral atoms, Science \textbf{394}, 197 (2026).


\bibitem{Papic2018}
C. J. Turner, A. A. Michailidis, D. A. Abanin, M. Serbyn, and Z. Papi\'{c}, Weak ergodicity breaking from quantum many-body scars, Nat. Phys. \textbf{14}, 745 (2018).

\bibitem{Lukinscars2019}
W. W. Ho, S. Choi, H. Pichler, and M. D. Lukin, Periodic orbits, entanglement, and quantum many-body scars in constrained models: Matrix product state approach, Phys. Rev. Lett. \textbf{122}, 040603 (2019).

\bibitem{Papic2021}
M. Serbyn, D. A. Abanin, and Z. Papi\'{c}, Quantum many-body scars and weak breaking of ergodicity, Nat. Phys. \textbf{17}, 675 (2021).

\bibitem{YCheng2024}
Y. Cheng and H. Zhai, Emergent U(1) lattice gauge theory in Rydberg atom arrays, Nat. Rev. Phys. \textbf{6}, 566 (2024).

\bibitem{Lukin2017}
H. Bernien, S. Schwartz, A. Keesling, H. Levine, A. Om-ran, H. Pichler, S. Choi, A. S. Zibrov, M. Endres, M. Greiner, V. Vuleti\'{c}, and M. D. Lukin, Probing many-body dynamics on a 51-atom quantum simulator, Nature \textbf{551}, 579 (2017).

\bibitem{Lukin2019}
A. Keesling, A. Omran, H. Levine, H. Bernien, H. Pichler, S. Choi, R. Samajdar, S. Schwartz, P. Silvi, S. Sachdev, P. Zoller, M. Endres, M. Greiner, V. Vuleti\'{c}, and M. D. Lukin, Quantum Kibble–Zurek mechanism and critical dynamics on a programmable Rydberg simulator,
Nature \textbf{568}, 207 (2019).

\bibitem{Dalmonte2020}
F. M. Surace, P. P. Mazza, G. Giudici, A. Lerose, A. Gambassi, and M. Dalmonte, Lattice Gauge Theories and String Dynamics in Rydberg Atom Quantum Simulators, Phys. Rev. X \textbf{10}, 021041 (2020).

\bibitem{LGT2025}
D. Gonz\'{a}lez-Cuadra et al., Observation of string breaking on a (2 + 1)D Rydberg quantum simulator, Nature \textbf{642}, 321 (2025).

\bibitem{LukinQSL2021}
G. Semeghini, H. Levine, A. Keesling, S. Ebadi, T. T. Wang, D. Bluvstein, R. Verresen, H. Pichler, M. Kalinowski, R. Samajdar, A. Omran, S. Sachdev, A. Vishwanath, M. Greiner, V. Vuleti\'{c}, and M. D. Lukin, Probing topological spin liquids on a programmable quantum simulator, Science \textbf{374}, 1242 (2021).

\bibitem{Browaeys2025}
M. Qiao, G. Emperauger, C. Chen, L. Homeier, S. Hollerith, G. Bornet, R. Martin, B. G\'{e}ly, L. Klein, D. Barredo, S. Geier, N.-C. Chiu, F. Grusdt, A. Bohrdt, T. Lahaye, and A. Browaeys, Realization of a doped quantum antiferromagnet in a Rydberg tweezer array, Nature \textbf{644}, 889 (2025).





\bibitem{Lan2016}
Z. Lan, W. Li, and I. Lesanovsky, Quantum melting of two-component Rydberg crystals, Phys. Rev. A  \textbf{94}, 051603(R) (2016).

\bibitem{Alicea2021}
K. Slagle, D. Aasen, H. Pichler, R. S. K. Mong, P. Fendley, X. Chen, M. Endres, and J. Alicea, Microscopic characterization of Ising conformal field theory in Rydberg chains, Phys. Rev. B \textbf{104}, 235109 (2021).

\bibitem{SUSY2024}
C. Li, S. Liu, H. Wang, W. Zhang, Z.-X. Li, H. Zhai, and Y. Gu, Uncovering Emergent Spacetime Supersymmetry with Rydberg Atom Arrays, Phys. Rev. Lett. \textbf{133}, 223401 (2024).

\bibitem{SUSY2025}
H. Wang, X. Li, C. Li, Tricritical Kibble-Zurek Scaling in Rydberg Atom Ladders, Nat. Commun. \textbf{16}, 10584 (2025).


\bibitem{Liu2026}
L.-Y.-N. Liu, S.-Y. Yu, S.-R. Peng, J. Sheng, S. Yi, P. Xu, S.-S. Gong, T. Shi, and J. Cui, Distinct many-body scars and emergent quantum phases driven by competing interactions in two-species Rydberg arrays, Phys. Rev. B \textbf{114}, 115406 (2026).

\bibitem{Saffman2008}
T. G. Walker and M. Saffman, Consequences of Zeeman degeneracy for the van der Waals blockade between Rydberg atoms, Phys. Rev. A \textbf{77}, 032723 (2008).
\bibitem{Cano2012}
D. Cano and J. Fort\'agh, Nonadditive potentials between three Rydberg atoms, Phys. Rev. A \textbf{86}, 043422 (2012).

\bibitem{XFShi2014}
X.-F. Shi, F. Bariani, and T. A. B. Kennedy, Entanglement of neutral-atom chains by spin-exchange Rydberg interaction, Phys. Rev. A \textbf{90}, 062327 (2014).

\bibitem{Beterov2015}
I. I. Beterov and M. Saffman, Rydberg blockade, F\"{o}rster resonances, and quantum state measurements with different atomic species, Phys. Rev. A \textbf{92}, 042710 (2015).

\bibitem{Mandoki2016}
A. Paris-Mandoki, H. Gorniaczyk, C. Tresp, I. Mirgorodskiy and S. Hofferberth, Tailoring Rydberg interactions via F\"{o}rster resonances: state combinations, hopping and angular dependence, J. Phys. B: At. Mol. Opt. Phys. \textbf{49}, 164001 (2016).

\bibitem{CSAdams2025}
K. Wadenpfuhl and C. S. Adams, Unraveling the structures in the van der Waals interactions of alkali-metal Rydberg atoms, Phys. Rev. A \textbf{111}, 062803 (2025).

\bibitem{LiZhai2026}
C. Li and H. Zhai, Universal Spin-Position Coupled Rydberg Interactions, arXiv: 2608.23694 (2026).



\bibitem{Sheng2022}
C. Sheng, J.-Y. Hou, X.-D. He, K.-P. Wang, R.-J. Guo, J. Zhuang, B. Mamat, P. Xu, M. Liu, J. Wang, and M.-S. Zhan, Defect-Free Arbitrary-Geometry Assembly of Mixed-Species Atom Arrays, Phys. Rev. Lett. \textbf{128}, 083202 (2022).

\bibitem{Bernien2022}
K. Singh, S. Anand, A. Pocklington, J. T. Kemp, and H. Bernien, Dual-Element, Two-Dimensional Atom Array with Continuous-Mode Operation, Phys. Rev. X \textbf{12}, 011040 (2022).

\bibitem{KKNi2022}
J. T. Zhang, L. R. B. Picard, W. B. Cairncross, K. Wang, Y. Yu, F. Fang, and K.-K. Ni, An optical tweezer array of ground-state polar molecules, Quantum Sci. Technol. \textbf{7}, 035006 (2022).

\bibitem{Takahashi2025}
Y. Nakamura, T. Kusano, R. Yokoyama, K. Saito, K. Higashi, N. Ozawa, T. Takano, Y. Takasu, and Y. Takahashi, Hybrid Atom Tweezer Array of Nuclear Spin and Optical Clock Qubits, Phys. Rev. X \textbf{14}, 041062 (2024).


\bibitem{Saffman2025}
C. Fang et al., Interleaved dual-species arrays of single atoms using a passive optical element and one trapping laser, Sci. Adv. \textbf{11}, eadw4166 (2025).

\bibitem{Anand2024} S. Anand, C. E. Bradley, R. White, V. Ramesh, K. Singh, and H. Bernien, A dual-species Rydberg array, Nat. Phys. \textbf{20}, 1744 (2024).

\bibitem{SaffmanRbCs2026}
J. Miles \textit{et al}, Qubit syndrome measurements with a high fidelity Rb-Cs Rydberg gate, arXiv: 2603.13492.

\bibitem{KKNi2026}
Yu Wang, Ryan Cimmino, Kenneth Wang, Santiago Lopez, Jeffrey Li, Jin Ming Koh, Jonathan N. Hallén, Anne Matthies, Norman Y. Yao, Kang-Kuen Ni, Multi-Qubit Stabilizer Readout on a Dual-Species Rydberg Array, arXiv: 2605.10924 (2026).

\bibitem{Chepiga2024} N. Chepiga, Tunable quantum criticality in multicomponent Rydberg arrays, Phys. Rev. Lett.  \textbf{132}, 076505 (2024).

\bibitem{Tomza2025}
J. Dobrzyniecki, P. Heim, and M. Tomza, Tunable two-species spin models with Rydberg atoms in circular and elliptical states, Phys. Rev. Res. \textbf{7}, 013321 (2025).

\bibitem{Pichler2026}
F. Cesa, A. D. Fini, D. A. Korbany, R. Tricarico, H. Bernien, H. Pichler, and L. Piroli, Engineering discrete local dynamics in globally driven dual-species atom arrays, arXiv: 2601.16961 (2026).

\bibitem{Browaeys2025exchange}
G. Emperauger, M. Qiao, G. Bornet, C. Chen, R. Martin, Y. T. Chew, B. Gély, L. Klein, D. Barredo, A. Browaeys, and T. Lahaye, Benchmarking direct and indirect dipolar spin-exchange interactions between two Rydberg atoms, Phys. Rev. A \textbf{111}, 062806 (2025).

\bibitem{CChen2026}
L. Ruan, Z. Zhou, Y. Guo, C. Li, and C. Chen, Observation of electron spin interactions between Rydberg atoms, arXiv: 2608.23711 (2026).

\bibitem{setup2019}
K.-P. Wang \textit{et al.}, Preparation of a heteronuclear two-atom system in the three-dimensional ground state in an optical tweezer, Phys. Rev. A \textbf{100}, 063429 (2019).

\bibitem{array2026}
M.-R. Wei \textit{et al.}, Enhanced atom-by-atom assembly of defect-free two-dimensional mixed-species atomic arrays, Phys. Rev. Applied \textbf{25}, 034009 (2026).

\bibitem{Browaeys2009}
A. Ga\"etan, Y. Miroshnychenko, T. Wilk, A. Chotia, M. Viteau, D. Comparat, P. Pillet, A. Browaeys, and P. Grangier, Observation of collective excitation of two individual atoms in the Rydberg blockade regime, Nat. Phys. \textbf{5}, 115 (2009).


\bibitem{Pairint2026}
J. M\"ogerle, F. Hummel, A. Keil, T. Legrand, E. J. Braun, H. Menke, J. King, B. Olmos, S. Hofferberth, H. P. Büchler, and S. Weber, The PairInteraction toolkit for modeling Rydberg physics in alkali and alkaline-earth-like atoms, Sci. Rep. \textbf{16}, 31484 (2026).

\bibitem{ARC}
N. \v{S}ibali\'c, J. D. Pritchard, K. J. Weatherill, C. S. Adams, ARC: An open-source library for calculating properties of alkali Rydberg atoms, Comput. Phys. Commun. \textbf{220}, 319 (2017)


\bibitem{Weidemuller2021}
A. Signoles, T. Franz, R. F. Alves, M. G\"arttner, S. Whitlock, G. Z\"urn, and M. Weidem\"uller, Glassy Dynamics in a Disordered Heisenberg Quantum Spin System, Phys. Rev. X \textbf{11}, 011011 (2021).


\end{thebibliography}
\end{document}